\documentclass[aps,prl,showpacs,twocolumn,superscriptaddress,floatfix]{revtex4}

\usepackage{graphicx}
\usepackage{amsmath}
\usepackage{amssymb}
\usepackage{bm}
\usepackage{dcolumn}

\begin{document}

\title{Orientation-Dependent Transparency of Metallic Interfaces}
\author{P. X. Xu}
\affiliation{State Key Laboratory for Surface Physics,
Institute of Physics,
Chinese Academy of Sciences,
P. O. Box 603,
Beijing 100080,
China.}
\author{K. Xia}
\affiliation{State Key Laboratory for Surface Physics,
Institute of Physics,
Chinese Academy of Sciences,
P. O. Box 603,
Beijing 100080,
China.}
\author{M. Zwierzycki}
\affiliation{Faculty of Science and Technology and MESA$^+$
Institute for Nanotechnology, University of Twente, P.O. Box 217,
7500 AE Enschede, The Netherlands}
\affiliation{Max-Planck-Institut f\"{u}r Festk\"{o}rperforschung,
Heisenbergstr. 1, D-70569 Stuttgart, Germany.}
%
%
\author{M. Talanana}
\affiliation{Faculty of Science and Technology and MESA$^+$
Institute for Nanotechnology, University of Twente, P.O. Box 217,
7500 AE Enschede, The Netherlands}
\author{P. J. Kelly}
\affiliation{Faculty of Science and Technology and MESA$^+$
Institute for Nanotechnology, University of Twente, P.O. Box 217,
7500 AE Enschede, The Netherlands}

\date{\today }

\begin{abstract}
As devices are reduced in size, interfaces start to dominate electrical
transport making it essential to be able to describe reliably how they
transmit and reflect electrons. For a number of nearly perfectly
lattice-matched materials, we calculate from first-principles the
dependence of the interface transparency on the crystal orientation.
Quite remarkably, the largest anisotropy is predicted for interfaces
between the prototype free-electron materials silver and aluminium for
which a massive factor of two difference between (111) and (001)
interfaces is found.
\end{abstract}


\maketitle


A recurring theme in condensed matter physics in the last twenty years
has been the discovery of new physical effects and properties in
systems with reduced dimensions; the prospect of exploiting these
effects and properties in logical processing, sensing and storage
devices is an important driving force behind nano-science and
 -technology. In semiconductors, the electronic structures of the
electrons responsible for conduction can be described using simple
models. The same is not true of the ferromagnetic transition metals
which form the basis for magnetoelectronics. It is the non-trivial
spin-dependence of the transmission and reflection of electrons at
magnetic interfaces which provides the key to understanding phenomena
such as oscillatory exchange coupling, giant- and tunneling-
magnetoresistance, spin transfer torque, spin-pumping and spin
injection \cite{UMS}. For well-studied material combinations such as
Co/Cu and Fe/Cr, modest spin-dependence of the interface transmission
\cite{Schep:prb97,Stiles:prb00,Xia:prb01} of the order of 10-20\% is
sufficient to account for experimental observations \cite{Bass:jmmm99}.

However, the confrontation of theory and experiment just referred
to is at best indirect and model-dependent. Even though the theory
of transport in small structures is formulated in terms of
transmission and reflection matrices \cite{Beenakker:rmp97},
measuring interface transparencies directly has proven quite
difficult \cite{Galinon:apl05}. To identify interfaces suitable for
experimental study, we have undertaken a systematic
materials-specific study of the orientation dependence of the
interface transmission between pairs of isostructural metals whose
lattice constants match within a percent or so in the hope that it
will prove possible to grow such interfaces epitaxially.

One of the metal pairs we studied was Al/Ag. Both metals have the fcc
crystal structure and their lattice constants are matched within 1\%.
Aluminium is a textbook \cite{Ashcroft:76} example of a system well
described by the (nearly) free electron model. Silver, also usually
assumed to be a free electron-like material, is a noble metal with high
conductivity which is frequently used for electrical contacting. We
found that in spite of the simplicity of both metals' electronic
structures the transmission through Al/Ag interfaces can differ quite
significantly from the predictions of the free electron model.
In particular, between (111) and (001) orientations we find a factor 2
difference in interface transmission for clean Al/Ag interfaces. For
free electrons the anisotropy should vanish. Our result is insensitive
to interface disorder. We identify a new factor responsible for this
difference which is not related to the standard velocity-
\cite{Mazin:prl99,Mazin:jap01} or symmetry-mismatch
\cite{Zwierzycki:prb03,Xia:prb06} mechanisms.

A free electron description of interface scattering, in which the
effect of the crystal potential on transport is completely
neglected, underlies the BTK theory \cite{Blonder:prb82} used to
interpret \cite{Mazin:prl99,Mazin:jap01} Andreev reflection (AR)
experiments. Point contact AR experiments are increasingly used to
identify the pairing symmetry of superconductors and, in the field
of magnetoelectronics, to determine the polarization of magnetic
materials \cite{Soulen:sc98,Upadhyay:prl98}. Our finding that the
electronic structure can have such a large effect on the interface
transmission, implies that experiments should be analysed using
more sophisticated models.

\begin{table}[b]
\begin{ruledtabular}
\caption[Tab1]{Sharvin conductances and interface transmissions in
units of $10^{15} \Omega^{-1}m^{-2}$, interface resistances $SR$
\cite{Schep:prb97,Xia:prb01} for ideal (and, in brackets, for
disordered) interfaces in units of $10^{-15} \Omega m^2$. $S$ is the
area of the sample for which R is measured. Interface disorder was
modelled in $ 10\times10$ lateral supercells with two layers of 50-50
alloy. The largest uncertainty between different configurations of
disorder is about 2.3\%. The values given are for a single spin. For
the pairs of materials and orientations indicated by a (*), comparison
of the interface resistances shown in the last two columns with
experimental values extracted from measurement on multilayers by the
MSU group \cite{Bass:jmmm99,Galinon:apl05} yields reasonable
quantitative agreement \cite{Schep:prb97,Xia:prb01,Xia:prb06}.}
\begin{tabular}{llcccc}
  $A/B$            &        & $G_A$ & $G_B$   & $G_{A/B}$      &  $2SR$        \\
\hline \hline
Al/Ag              & (111)  & 0.69  &  0.45   &  0.41 (0.36)   &  0.64 (0.92)  \\
a$_{fcc}=4.05$ \AA & (110)  & 0.68  &  0.47   &  0.30 (0.32)   &  1.60 (1.39)  \\
                   & (001)  & 0.73  &  0.45   &  0.22 (0.24)   &  2.82 (2.37)  \\
\hline
Al/Au              & (111)  & 0.69  &  0.44   &  0.41 (0.35)   &  0.60 (0.99)  \\
a$_{fcc}=4.05$ \AA & (001)  & 0.73  &  0.46   &  0.24 (0.26)   &  2.37 (2.14)  \\
\hline
Pd/Pt              & (111)  & 0.62  &  0.71   &  0.55 (0.54)   &  0.30 (0.33)  \\
a$_{fcc}=3.89$ \AA & (001)  & 0.58  &  0.70   &  0.52 (0.51)   &  0.37 (0.39)  \\
\hline
W/Mo               & (001)  & 0.45  &  0.59   &  0.42 (0.42)   &  0.42 (0.42)  \\
a$_{bcc}=3.16$ \AA & (110)  & 0.40  &  0.54   &  0.37 (0.38)   &  0.52 (0.47)  \\
\hline Cu/Co       & (111)* & 0.56  &  0.47   &  0.43 (0.43)   &  0.34 (0.35)  \\
majority           & (001)  & 0.55  &  0.49   &  0.46 (0.45)   &  0.26 (0.27)  \\
a$_{fcc}=3.61$ \AA & (110)  & 0.59  &  0.50   &  0.46 (0.46)   &  0.35 (0.35)  \\
\hline
Cu/Co              & (111)* & 0.56  &  1.05   &  0.36 (0.31)   &  1.38 (1.82)  \\
minority           & (001)  & 0.55  &  1.11   &  0.32 (0.32)   &  1.79 (1.79)  \\
a$_{fcc}=3.61$ \AA & (110)  & 0.59  &  1.04   &  0.31 (0.35)   &  1.89 (1.55)  \\
\hline
Cr/Fe              & (111)  & 0.61  &  0.82   &  0.27 (0.31)   &  2.22 (1.84)  \\
majority           & (001)  & 0.64  &  0.82   &  0.11 (0.25)   &  7.46 (2.55)  \\
a$_{bcc}=2.87$ \AA & (110)* & 0.59  &  0.78   &  0.22 (0.27)   &  3.04 (2.18)  \\
\hline
Cr/Fe              & (111)  & 0.61  &  0.41   &  0.34 (0.34)   &  0.93 (0.95)  \\
minority           & (001)  & 0.64  &  0.46   &  0.35 (0.35)   &  0.98 (0.95)  \\
a$_{bcc}=2.87$ \AA & (110)* & 0.59  &  0.40   &  0.32 (0.32)   &  1.03 (1.06)  \\
\end{tabular}
\label{tableone}
\end{ruledtabular}
\end{table}

Our study was based upon first-principles calculations of the
interface electronic structure performed within the framework of
density functional theory (DFT) and the local spin density
approximation (LSDA). Bulk and interface potentials were
determined self-consistently using the tight binding linearized
muffin tin orbital (TB-LMTO) \cite{Andersen:85} surface Green's
function method \cite{Turek:97}. We assumed common lattice
constants for both metals of a given structure e.g.
$a_{Al}=a_{Ag}=4.05$ \AA . The potentials obtained in this way
were used as input to a TB-MTO wavefunction-matching
\cite{Xia:prb01,Xia:prb06} calculation of the transmission and
reflection coefficients between Bloch states on either side of the
interface. The efficiency of this approach is such that interface
disorder can be modelled using large lateral supercells. For
disordered systems, the potentials were calculated using the layer
CPA (coherent potential approximation). The results of the
calculations for a number of lattice-matched materials are
summarized in Table~\ref{tableone}.

The Sharvin conductances, $G_A$ and $G_B$, reported in the third
and fourth columns of Table~\ref{tableone} are proportional to the
number of states at the Fermi level propagating in the transport
direction. They are properties of the bulk materials which are
determined by the area of the Fermi surface projections and are a
measure of the current-carrying capacity of the conductor in the
ballistic regime. The largest intrinsic orientation dependence,
seen to be about 13 \%, is found for Mo; for Al and Ag,
respectively, it is less than 8\% and 5\%.

The interface transmission in column five of Table~\ref{tableone} is
expressed as a conductance, $G_{A/B}=e^2/h \sum_{\mu\nu}T_{\mu\nu}$,
where $T_{\mu\nu}$ is the probability for the incoming state $\nu$ in
material A to be transmitted through the interface into the outgoing
state $\mu$ in material B. For most pairs of materials \cite{note3},
the orientation dependence of $G_{A/B}$ is modest ($\sim 15 \%$ for
Mo$/$W) and the interface conductance itself tends to be slightly
smaller than the lower of the two Sharvin conductances. For these
systems the behaviour of the transmission appears to be determined by
the projection of the Fermi surfaces. However, this is not so for Al/Ag
and Al/Au interfaces. Here we observe a large anisotropy in the
transport properties. The factor 2 difference in transmission between
(111) and (001) orientations \cite{note4} results in a factor 4
difference between interface resistances estimated using the method of
\cite{Schep:prb97,Xia:prb01}.

\begin{figure}[t]
\includegraphics[scale=0.45,clip=true,bb=15 150 582 680]{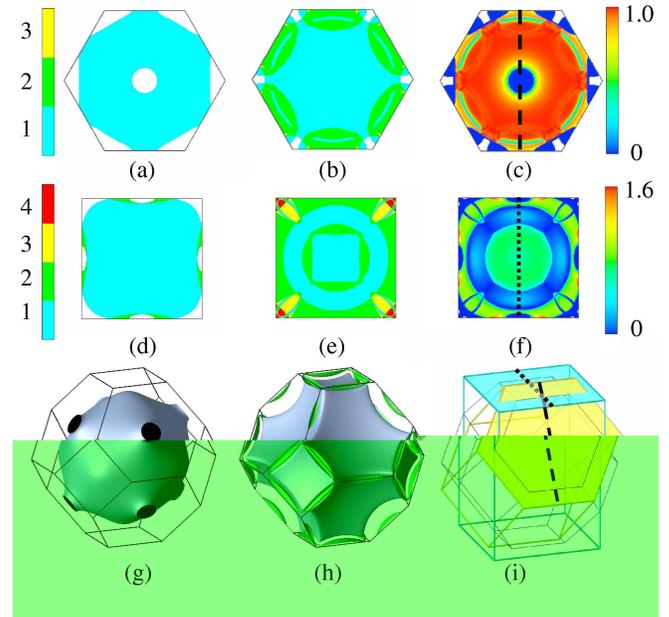}
\caption{
  Top row: Fermi surface projections for (a) Ag, (b) Al and (c)
  transmission probabilities in 2DBZ for (111) orientation.
  Middle row: Same for (001) orientation. The colour bars on the
  left indicate the number of scattering states in the leads for a
  given two dimensional wavevector ${\bf k}_{\parallel}$.
  The transmission probabilities indicated by the colour bars on the
  right can exceed 1 for ${\bf k}_{\parallel}$s for which there is more
  than one scattering state in both Ag and Al.
  Bottom row: Fermi surfaces of (g) Ag and (h) Al, (i) the interface
  adapted BZ for (001) and (111) orientations.
  The vertical dashed line in (c) and on the yellow plane
  in panel (i) indicate the cross-section used in the left-hand panel
  of Fig.~\ref{fig:cross} while
  the vertical dotted line in (f) and on the blue plane in
  panel (i) indicate the cross-section used in the right-hand panel.}
\label{proall}
\end{figure}

The transmission probability for the (111) and (001) orientations
is plotted in Figs.~\ref{proall}c and \ref{proall}f as a function
of the wave-vector component parallel to the interface, ${\bf
k}_{\parallel}$, within the 2D interface Brillouin zones (BZ). A
qualitative difference between the two orientations can be
observed. In the (111) case, the transmission is almost uniformly
high wherever there are states on both sides of the interface. The
(001) orientation exhibits more variation with high transmission
in the central and outer regions of the 2D~BZ but much lower in a
ring-shaped region in between. The presence of this ``cold ring''
is the reason why the total transmission is lower for the (001)
orientation. Explaining the transparency anisotropy of Al/Ag
interfaces requires finding an explanation for the low
transmission values in this region of the 2D~BZ.

Two mechanisms are usually taken into account when analysing the
scattering at perfect interfaces. The first, velocity mismatch, is
the modulation of the transmission by a factor reminiscent of the
free electron formula for the transmission through a potential
step: $T=4v_L v_R/(v_L+ v_R)^2$ where $v_{L/R}$ are the components
of the Fermi velocities in the transport direction on the left and
right sides of the interface. This modulation is indeed present in
our calculated transmissions but its effect tends to be noticeable
only when one of the velocities is almost vanishingly small. Naive
application of the free electron formula yields uniformly good
transmission \cite{note1} independent of the orientation.
Symmetry mismatch, the second mechanism, can suppress the transmission
between states of incompatible symmetries (\emph{e.g.} even vs. odd
etc.). Examination of the eigenvectors demonstrates that this is not
the case for the Al/Ag system. For example, states on both sides of the
interface, with ${\bf k}_{\parallel}$ along the vertical dotted line in
Fig.~\ref{proall}f, are even under reflection in the plane defined by
this line and the (001) transport direction. Their orbital composition
($s$,$p_y$,$p_z$,$d_{yz}$,$d_{3z^2-r^2}$,$d_{x^2-y^2}$ where the $y$
axis is parallel to the dotted line and $z$ is the transport direction)
is essentially the same for both materials. The same holds for states
along other symmetry lines/planes and general ${\bf k}_{\parallel}$
points (in the sense of orbital composition). The origin of the ``cold
ring'' must be sought elsewhere.

\begin{figure}[t]
\includegraphics[scale=0.18]{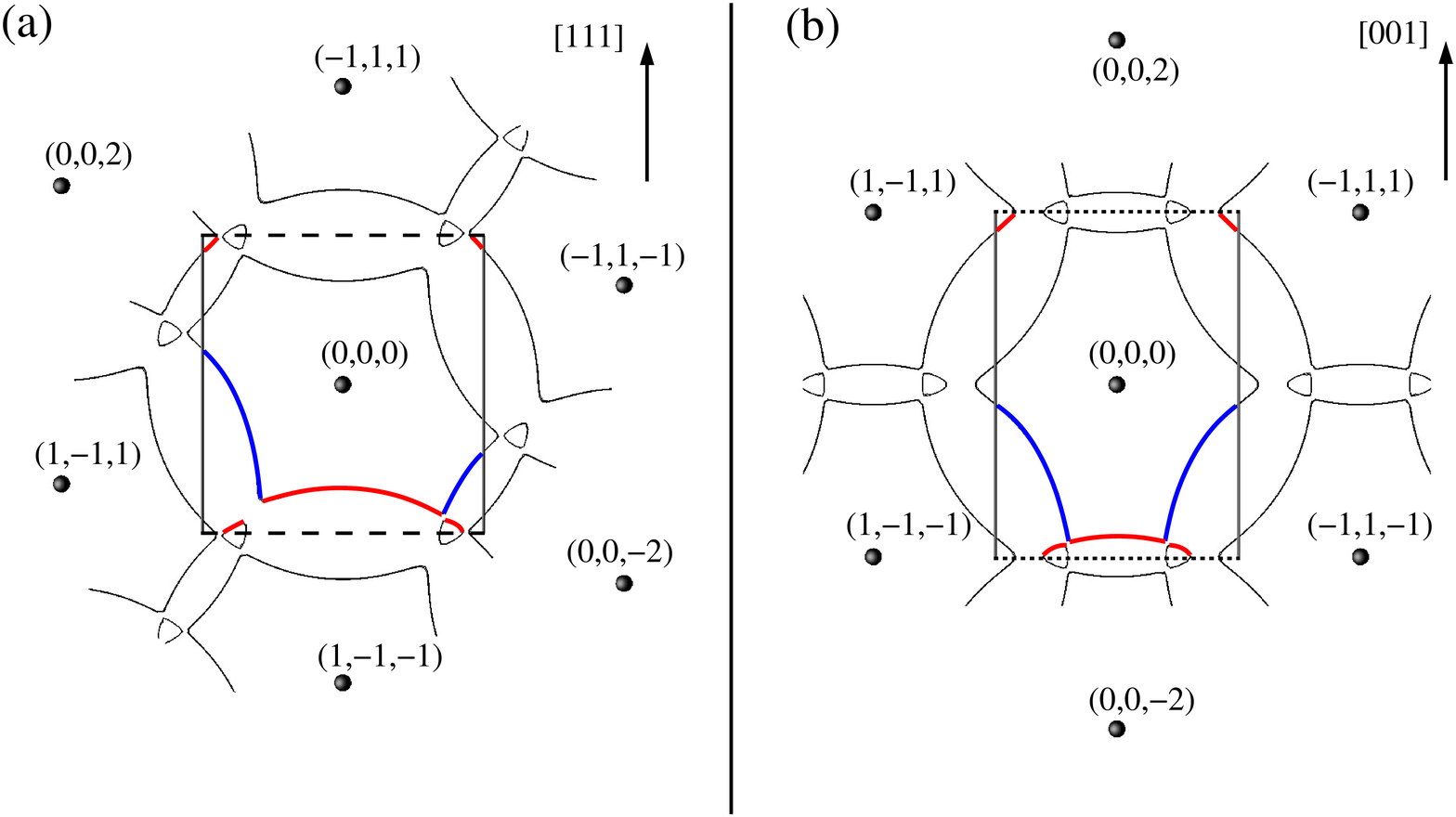}
\caption{Intersection of a (110) plane with the Al Fermi surface
  and with the interface adapted BZs indicated in Fig.~\ref{proall}i
  (where the meaning of the dashed and dotted lines is explained). The
  labelled dots indicate the positions of the RL sites with coordinates
  given in units of $2\pi/a$.  The red (blue) lines indicate regions of
  high (low) transmission.}
\label{fig:cross}
\end{figure}

In spite of the failure of the free electron transmission formula,
this simple model serves as a useful starting point for analysing the
Fermi surface (FS) topologies. In the simplest possible approach, we
model the FS of Ag (shown in Fig.~\ref{proall}g) as a sphere which
fits into the first BZ. A larger sphere, accommodating three
electrons, is needed for trivalent Al. In an extended zone scheme,
conservation of momentum parallel to the interface dictates that the
transmission through a specular interface is non-zero only between
states with the same values of ${\bf k}_{\parallel}$; these are the
${\bf k}_{\parallel}$-vectors belonging to the region where
projections of the Fermi spheres on a plane perpendicular to the
transport direction overlap. For systems with lattice periodicity, we
must use a downfolded FS, with fragments of the original FS sphere
back-translated into the 1st Brillouin zone, a procedure which can be
realized geometrically by placing spheres accommodating three
electrons on reciprocal lattice (RL) sites and then only considering
the fragments in the first BZ. Examination of the FS of Al calculated
from first-principles (Fig.~\ref{proall}h) and its cross-section
(Fig.~\ref{fig:cross}) reveals that, in spite of its apparent
complexity, it remains essentially (piecewise) spherical. For some
values of $\mathbf{k}_{||}$ (see Figs.~\ref{proall}b and
\ref{proall}e), Al can now have more than one propagating state.
Nevertheless, in the free electron limit, the downfolded states are
strictly orthogonal to the states in Ag and the total transmission is
unchanged. For a reduced zone scheme, we formulate the following rule:
\emph{The transmission between states in two (nearly) free electron
  materials which have the same ${\bf k}_{\parallel}$, but originate
  from reciprocal lattice sites whose parallel components \emph{do
    not} coincide, vanishes in the free electron limit and is expected
  to be strongly suppressed for nearly free electron like materials.}

Obviously, the truly free electron system can not exhibit anisotropy.
However, in the presence of the periodic potential the original,
piecewise-spherical Fermi surface and consequently the transmission
is going to be modified. Firstly, since the wave functions are no
longer pure plane waves, the strict orthogonality of the downfolded
states is relaxed and the transmission can assume finite although
typically small values (hence \emph{suppressed} instead of \emph{zero}
in the above rule).  Secondly, the shape of the Fermi surface changes
with the modifications being strongest in the vicinity of RL planes
where, for Al, we observe the opening of gaps between previously
connected fragments. The anisotropy is mostly related to this second
effect.

In Fig.~\ref{fig:cross}, we show the intersection of the Al FS with a
(110) plane. The two plots are rotated so that the vertical axis in
Fig.~\ref{fig:cross}a is the [111] direction while in
Fig.~\ref{fig:cross}b it is [001]. In both cases the positions of the
nearest RL sites (on which spheres are centred) are shown together
with the cross-section through the relevant interface-adapted
Brillouin zone, which is different for each orientation; see
Fig.~\ref{proall}i. We can now readily identify spheres from which
various fragments of the Fermi surface originate and mark those
fragments with positive (upward) velocities, according to the rule
given above, as having high (red) or low (blue) transmissions. In the
(001) case, the ``high'' fragments originate from (0,0,0) and (0,0,-2)
centred spheres. Comparing Figs.~\ref{proall}f and \ref{fig:cross}b,
we note that the position of the gaps opened between these spheres by
Bragg reflection on the $(001)$ and $(00\bar1)$ planes coincides, in
projection along the [001] direction, with the position of the ``cold
ring'' in Fig.~\ref{proall}f. The other states present in this region
originate from (1,-1,-1) (and equivalent) centred spheres, are
therefore nearly orthogonal to states in Ag centered on (0,0,0) and
so have low transmission. In the (111) case however, the large
fragments of FS belonging to the same (1,-1,-1) sphere have high
transmissions (Fig.~\ref{fig:cross}a) and dominate transport.  In
addition, the effect of gap-opening is reduced in this orientation
because of the rotation. Combination of these two factors results in
the almost uniformly high transmission seen in Fig.~\ref{proall}c.

We can now finally identify the origin of the transmission anisotropy
for Al/Ag interface. It stems from two factors: (i) the near
orthogonality of the downfolded Al states to those belonging to the
simple Ag sphere and (ii) the gaps opened in the continuous free
electron Fermi surface by the periodic potential. The latter factor is
of course related to the symmetry of the underlying crystal lattice and
directly responsible for the introduction of the orientation
dependence. For Al/Au interfaces, the interface transmissions and
resistances are very similar to the Al/Ag case.

The orientation dependence of the interface transmission of six metal
pairs with the same structure and lattice parameter was calculated.
For \textit{fcc} Ag/Al a factor two difference between the (111) and
(001) orientations was found and explained within the free electron
model. The predicted anisotropic interface resistance and Andreev
reflection (not shown) are not very sensitive to interface disorder
and should be observable experimentally.

\emph{Acknowledgments :}This work is part of the research program of
the ``Stichting voor Fundamenteel Onderzoek der Materie'' (FOM) and
the use of supercomputer facilities was sponsored by the ``Stichting
Nationale Computer Faciliteiten'' (NCF), both financially supported by
the ``Nederlandse Organisatie voor Wetenschappelijk Onderzoek'' (NWO).
It is also supported by ``NanoNed'', a nanotechnology programme of the
Dutch Ministry of Economic Affairs.
We are grateful to Anton Starikov for permission to use his sparse
matrix version of the TB-MTO code.


\begin{thebibliography}{20}
\expandafter\ifx\csname
natexlab\endcsname\relax\def\natexlab#1{#1}\fi
\expandafter\ifx\csname bibnamefont\endcsname\relax
  \def\bibnamefont#1{#1}\fi
\expandafter\ifx\csname bibfnamefont\endcsname\relax
  \def\bibfnamefont#1{#1}\fi
\expandafter\ifx\csname citenamefont\endcsname\relax
  \def\citenamefont#1{#1}\fi
\expandafter\ifx\csname url\endcsname\relax
  \def\url#1{\texttt{#1}}\fi
\expandafter\ifx\csname
urlprefix\endcsname\relax\def\urlprefix{URL }\fi
\providecommand{\bibinfo}[2]{#2}
\providecommand{\eprint}[2][]{\url{#2}}

\bibitem[{\citenamefont{see the collection~of articles}(1994-2005)}]{UMS}
\bibinfo{author}{\bibnamefont{see the collection~of articles}}, in
  \emph{\bibinfo{booktitle}{Ultrathin Magnetic Structures I-IV}}, edited by
  \bibinfo{editor}{\bibfnamefont{J.~A.~C.} \bibnamefont{Bland}}
  \bibnamefont{and} \bibinfo{editor}{\bibfnamefont{B.}~\bibnamefont{Heinrich}}
  (\bibinfo{publisher}{Springer-Verlag}, \bibinfo{address}{Berlin},
  \bibinfo{year}{1994-2005}).

\bibitem[{\citenamefont{Schep et~al.}(1997)\citenamefont{Schep, van Hoof,
  Kelly, Bauer, and Inglesfield}}]{Schep:prb97}
\bibinfo{author}{\bibfnamefont{K.~M.} \bibnamefont{Schep}},
  \bibnamefont{et~al.},
  \bibinfo{journal}{Phys. Rev. B}
  \textbf{\bibinfo{volume}{56}}, \bibinfo{pages}{10805} (\bibinfo{year}{1997}).

\bibitem[{\citenamefont{Stiles and Penn}(2000)}]{Stiles:prb00}
\bibinfo{author}{\bibfnamefont{M.~D.} \bibnamefont{Stiles}} \bibnamefont{and}
  \bibinfo{author}{\bibfnamefont{D.~R.} \bibnamefont{Penn}},
  \bibinfo{journal}{Phys. Rev. B} \textbf{\bibinfo{volume}{61}},
  \bibinfo{pages}{3200} (\bibinfo{year}{2000}).

\bibitem[{\citenamefont{Xia et~al.}(2001)\citenamefont{Xia, Kelly, Bauer,
  Turek, Kudrnovsk\'{y}, and Drchal}}]{Xia:prb01}
\bibinfo{author}{\bibfnamefont{K.}~\bibnamefont{Xia}},
  \bibnamefont{et~al.},
  \bibinfo{journal}{Phys. Rev. B}
  \textbf{\bibinfo{volume}{63}},
  \bibinfo{pages}{064407} (\bibinfo{year}{2001}).

\bibitem[{\citenamefont{Bass and {Pratt Jr.}}(1999)}]{Bass:jmmm99}
\bibinfo{author}{\bibfnamefont{J.}~\bibnamefont{Bass}} \bibnamefont{and}
  \bibinfo{author}{\bibfnamefont{W.~P.} \bibnamefont{{Pratt Jr.}}},
  \bibinfo{journal}{J. Magn. \& Magn. Mater.} \textbf{\bibinfo{volume}{200}},
  \bibinfo{pages}{274} (\bibinfo{year}{1999}).

\bibitem[{\citenamefont{Beenakker}(1997)}]{Beenakker:rmp97}
\bibinfo{author}{\bibfnamefont{C.~W.~J.} \bibnamefont{Beenakker}},
  \bibinfo{journal}{Rev. Mod. Phys.} \textbf{\bibinfo{volume}{69}},
  \bibinfo{pages}{731} (\bibinfo{year}{1997}).

\bibitem[{\citenamefont{Galinon et~al.}(2005)\citenamefont{Galinon, Tewolde,
  Loloee, Chiang, Olson, Kurt, {Pratt Jr.}, Bass, Xu, Xia
  et~al.}}]{Galinon:apl05}
\bibinfo{author}{\bibfnamefont{C.}~\bibnamefont{Galinon}},
  \bibinfo{author}{\bibfnamefont{K.}~\bibnamefont{Tewolde}},
  \bibinfo{author}{\bibfnamefont{R.}~\bibnamefont{Loloee}},
  \bibinfo{author}{\bibfnamefont{W.~C.} \bibnamefont{Chiang}},
  \bibinfo{author}{\bibfnamefont{S.}~\bibnamefont{Olson}},
  \bibinfo{author}{\bibfnamefont{H.}~\bibnamefont{Kurt}},
  \bibinfo{author}{\bibfnamefont{W.~P.} \bibnamefont{{Pratt Jr.}}},
  \bibinfo{author}{\bibfnamefont{J.}~\bibnamefont{Bass}},
  \bibinfo{author}{\bibfnamefont{P.~X.} \bibnamefont{Xu}},
  \bibinfo{author}{\bibfnamefont{K.}~\bibnamefont{Xia}}, \bibnamefont{et~al.},
  \bibinfo{journal}{Appl. Phys. Lett.} \textbf{\bibinfo{volume}{86}},
  \bibinfo{pages}{182502} (\bibinfo{year}{2005}).

\bibitem[{\citenamefont{Ashcroft and Mermin}(1976)}]{Ashcroft:76}
\bibinfo{author}{\bibfnamefont{N.~W.} \bibnamefont{Ashcroft}} \bibnamefont{and}
  \bibinfo{author}{\bibfnamefont{N.~D.} \bibnamefont{Mermin}},
  \emph{\bibinfo{title}{Solid State Physics}}
  (\bibinfo{publisher}{Holt-Saunders International Editions},
  \bibinfo{address}{Philadelphia}, \bibinfo{year}{1976}).

\bibitem[{\citenamefont{Mazin}(1999)}]{Mazin:prl99}
\bibinfo{author}{\bibfnamefont{I.~I.} \bibnamefont{Mazin}},
  \bibinfo{journal}{Phys. Rev. Lett.} \textbf{\bibinfo{volume}{83}},
  \bibinfo{pages}{1427} (\bibinfo{year}{1999}).

\bibitem[{\citenamefont{Mazin et~al.}(2001)\citenamefont{Mazin, Golubov, and
  Nadgorny}}]{Mazin:jap01}
\bibinfo{author}{\bibfnamefont{I.~I.} \bibnamefont{Mazin}},
  \bibinfo{author}{\bibfnamefont{A.~A.} \bibnamefont{Golubov}},
  \bibnamefont{and} \bibinfo{author}{\bibfnamefont{B.}~\bibnamefont{Nadgorny}},
  \bibinfo{journal}{J. Appl. Phys.} \textbf{\bibinfo{volume}{89}},
  \bibinfo{pages}{7576} (\bibinfo{year}{2001}).

\bibitem[{\citenamefont{Zwierzycki et~al.}(2003)\citenamefont{Zwierzycki, Xia,
  Kelly, Bauer, and Turek}}]{Zwierzycki:prb03}
\bibinfo{author}{\bibfnamefont{M.}~\bibnamefont{Zwierzycki}},
  \bibnamefont{et~al.},
  \bibinfo{journal}{Phys. Rev. B} \textbf{\bibinfo{volume}{67}},
  \bibinfo{pages}{092401} (\bibinfo{year}{2003}).

\bibitem[{\citenamefont{Xia et~al.}(2006)\citenamefont{Xia, Zwierzycki,
  Talanana, Kelly, and Bauer}}]{Xia:prb06}
\bibinfo{author}{\bibfnamefont{K.}~\bibnamefont{Xia}},
  \bibnamefont{et~al.},
  \bibinfo{journal}{Phys. Rev. B} \textbf{\bibinfo{volume}{73}},
  \bibinfo{pages}{064420} (\bibinfo{year}{2006}).

\bibitem[{\citenamefont{Blonder et~al.}(1982)\citenamefont{Blonder, Tinkham,
  and Klapwijk}}]{Blonder:prb82}
\bibinfo{author}{\bibfnamefont{G.~E.} \bibnamefont{Blonder}},
  \bibinfo{author}{\bibfnamefont{M.}~\bibnamefont{Tinkham}}, \bibnamefont{and}
  \bibinfo{author}{\bibfnamefont{T.~M.} \bibnamefont{Klapwijk}},
  \bibinfo{journal}{Phys. Rev. B} \textbf{\bibinfo{volume}{25}},
  \bibinfo{pages}{4515} (\bibinfo{year}{1982}).

\bibitem[{\citenamefont{{Soulen Jr.} et~al.}(1998)\citenamefont{{Soulen Jr.},
  Byers, Osofsky, Nadgorny, Ambrose, Cheng, Broussard, Tanaka, Nowak, Moodera
  et~al.}}]{Soulen:sc98}
\bibinfo{author}{\bibfnamefont{R.~J.} \bibnamefont{{Soulen Jr.}}},
  \bibinfo{author}{\bibfnamefont{J.~M.} \bibnamefont{Byers}},
  \bibinfo{author}{\bibfnamefont{M.~S.} \bibnamefont{Osofsky}},
  \bibinfo{author}{\bibfnamefont{B.}~\bibnamefont{Nadgorny}},
  \bibinfo{author}{\bibfnamefont{T.}~\bibnamefont{Ambrose}},
  \bibinfo{author}{\bibfnamefont{S.~F.} \bibnamefont{Cheng}},
  \bibinfo{author}{\bibfnamefont{P.~R.} \bibnamefont{Broussard}},
  \bibinfo{author}{\bibfnamefont{C.~T.} \bibnamefont{Tanaka}},
  \bibinfo{author}{\bibfnamefont{J.}~\bibnamefont{Nowak}},
  \bibinfo{author}{\bibfnamefont{J.~S.} \bibnamefont{Moodera}},
  \bibnamefont{et~al.}, \bibinfo{journal}{Science}
  \textbf{\bibinfo{volume}{282}}, \bibinfo{pages}{85} (\bibinfo{year}{1998}).

\bibitem[{\citenamefont{Upadhyay et~al.}(1998)\citenamefont{Upadhyay,
  Palanisami, Louie, and Buhrman}}]{Upadhyay:prl98}
\bibinfo{author}{\bibfnamefont{S.~K.} \bibnamefont{Upadhyay}},
  \bibnamefont{et~al.},
  \bibinfo{journal}{Phys. Rev. Lett.} \textbf{\bibinfo{volume}{81}},
  \bibinfo{pages}{3247} (\bibinfo{year}{1998}).

\bibitem[{\citenamefont{Andersen et~al.}(1985)\citenamefont{Andersen, Jepsen,
  and Gl{\"{o}}tzel}}]{Andersen:85}
\bibinfo{author}{\bibfnamefont{O.~K.} \bibnamefont{Andersen}},
  \bibinfo{author}{\bibfnamefont{O.}~\bibnamefont{Jepsen}}, \bibnamefont{and}
  \bibinfo{author}{\bibfnamefont{D.}~\bibnamefont{Gl{\"{o}}tzel}}, in
  \emph{\bibinfo{booktitle}{Highlights of Condensed Matter Theory}}, edited by
  \bibinfo{editor}{\bibfnamefont{F.}~\bibnamefont{Bassani}},
  \bibinfo{editor}{\bibfnamefont{F.}~\bibnamefont{Fumi}}, \bibnamefont{and}
  \bibinfo{editor}{\bibfnamefont{M.~P.} \bibnamefont{Tosi}}
  (\bibinfo{publisher}{North-Holland}, \bibinfo{address}{Amsterdam},
  \bibinfo{year}{1985}), pp. \bibinfo{pages}{59--176}.

\bibitem[{\citenamefont{Turek et~al.}(1997)\citenamefont{Turek, Drchal,
  Kudrnovsk\'{y}, \v{S}ob, and Weinberger}}]{Turek:97}
\bibinfo{author}{\bibfnamefont{I.}~\bibnamefont{Turek}},
  \bibinfo{author}{\bibfnamefont{V.}~\bibnamefont{Drchal}},
  \bibinfo{author}{\bibfnamefont{J.}~\bibnamefont{Kudrnovsk\'{y}}},
  \bibinfo{author}{\bibfnamefont{M.}~\bibnamefont{\v{S}ob}}, \bibnamefont{and}
  \bibinfo{author}{\bibfnamefont{P.}~\bibnamefont{Weinberger}},
  \emph{\bibinfo{title}{Electronic Structure of Disordered Alloys, Surfaces and
  Interfaces}} (\bibinfo{publisher}{Kluwer},
  \bibinfo{address}{Boston-London-Dordrecht}, \bibinfo{year}{1997}).

\bibitem[{not({\natexlab{a}})}]{note3}
\bibinfo{note}{Fe/Cr is an exception. For the majority spin channel, a large
  orientation dependence of the interface tranmission is predicted. Unlike in
  the case of Al/Ag, this result is very sensitive to interface disorder. In
  addition, a single spin channel cannot be studied directly making it
  difficult to obtain an unambiguous experimental result.}

\bibitem[{not({\natexlab{b}})}]{note4}
\bibinfo{note}{We performed an extensive series of total energy calculations
  using LDA and GGA approximations to relax the various Al/Ag interfaces. Only
  a small dependence of the interface energy on the orientation was found. The
  transport calculations were repeated using the resulting relaxed geometries.
  The effect on the interface transmission is less than 3\% which is negligible
  on the scale of the predicted factor of two orientation dependence.}

\bibitem[{not({\natexlab{c}})}]{note1}
\bibinfo{note}{Morover the free electron formula would lead to the violation of
  the unitarity of the scattering matrix (\emph{i.e.} the conservation of
  particles) whenever there is more than one state on either side of the
  interface.}

\end{thebibliography}
\end{document}